\documentclass[a4paper]{aa}  

\usepackage{graphicx}
\usepackage{hyperref}
\usepackage{color}
\usepackage[varg]{txfonts}
\usepackage[T1]{fontenc}
\usepackage{ae,aecompl}
\usepackage[figure,figure*]{hypcap}
\usepackage{url}
\usepackage{amsmath}	
\usepackage{amssymb}	
\usepackage{tabularx}

\begin{document} 
\titlerunning{3mm SiO maser emission from YSO Orion Source~I}
\authorrunning{S.~Issaoun et al.}
   \title{VLBA imaging of the 3mm SiO maser emission in the disk-wind from the massive protostellar system Orion Source~I}


   \author{S.~Issaoun\inst{1} \and
          C.~Goddi\inst{1}$^{,}$\inst{2} \and
          L.~D.~Matthews\inst{3} \and
          L.~J.~Greenhill\inst{4} \and
          M.~D.~Gray\inst{5}\and
          E.~M.~L.~Humphreys\inst{6} \and
          C.~J.~Chandler\inst{7} \and
          M.~Krumholz\inst{8} \and
          H.~Falcke\inst{1}
          }

   \institute{Department of Astrophysics/IMAPP, Radboud University, PO Box 9010, 6500 GL Nijmegen, The Netherlands
         \and
             ALLEGRO/Leiden Observatory, Leiden University, PO Box 9513, NL-2300 RA Leiden, The Netherlands
         \and 
             MIT Haystack Observatory, Off Route 40, Westford, MA 01886, USA
         \and
			Harvard-Smithsonian Center for Astrophysics, 60 Garden Street, Cambridge, MA 02138, USA
         \and
			Jodrell Bank Centre for Astrophysics, Alan Turing Building, University of Manchester, Manchester M13 9PL, UK
	  	 \and 
         	European Southern Observatory, Karl-Schwarzschild-Strasse 2 D-85748 Garching bei M{\"u}nchen, Germany
         \and 
			National Radio Astronomy Observatory, P.O. Box O, Soccoro, NM 87801, USA
         \and 
            Research School of Astronomy \& Astrophysics, Australian National University, Canberra, ACT 2611, Australia
             }


 \abstract{High-mass star formation remains poorly understood due to observational difficulties (e.g. high dust extinction and large distances) hindering the resolution of disk-accretion and outflow-launching regions.}
{Orion Source~I is the closest known massive young stellar object (YSO) and exceptionally powers vibrationally-excited SiO masers at radii within 100\,AU, providing a unique probe of gas dynamics and energetics. We seek to observe and image these masers with Very Long Baseline Interferometry (VLBI).}
{We present the first images of the $^{28}$SiO $v=1$, $J=2-1$ maser emission around Orion Source~I observed at 86\,GHz ($\lambda$3mm) with the Very Long Baseline Array (VLBA). These images have high spatial ($\sim$0.3\,mas) and spectral ($\sim$0.054\,km\,s$^{-1}$) resolutions.}
{We find that the $\lambda$3mm masers lie in an X-shaped locus consisting of four arms, with blue-shifted emission in the south and east arms and red-shifted emission in the north and west arms. Comparisons with previous images of the $^{28}$SiO $v=1,2$, $J=1-0$ transitions at $\lambda$7mm (observed in 2001--2002) show that the bulk of the $J=2-1$ transition emission follows the streamlines of the $J=1-0$ emission and exhibits an overall velocity gradient consistent with the gradient at $\lambda$7mm. While there is spatial overlap between the $\lambda$3mm and $\lambda$7mm transitions, the $\lambda$3mm emission, on average, lies at larger projected distances from Source~I ($\sim$44\,AU compared with $\sim$35\,AU for $\lambda$7mm). The spatial overlap between the $v=1$, $J=1-0$ and $J=2-1$ transitions is suggestive of a range of temperatures and densities where physical conditions are favorable for both transitions of a same vibrational state. However, the observed spatial offset between the bulk of emission at $\lambda$3mm and $\lambda$7mm possibly indicates different ranges of temperatures and densities for optimal excitation of the masers. 
We discuss different maser pumping models that may explain the observed offset.} 
{We interpret the $\lambda$3mm and $\lambda$7mm masers as being part of a single wide-angle outflow arising from the surface of an edge-on disk rotating about a northeast-southwest axis, with a continuous velocity gradient indicative of differential rotation consistent with a Keplerian profile in a high-mass proto-binary.} 

   \maketitle

\section{Introduction}
Massive stars are crucial to the evolution of galaxies: they are the main source of heavy elements and UV radiation and they provide turbulence to the interstellar medium, greatly affecting star and planet formation processes \citep{zinn}. However, high-mass star formation still remains a poorly understood phenomenon due to observational difficulties, mainly high dust extinction, clustering and large distances of high-mass star-forming regions, which hinder attempts to resolve disk-accretion and outflow-launching regions \citep{zinn, beuther, beltran}. 

With very long baseline interferometry (VLBI), maser radiation from different molecules (OH, H$_2$O, CH$_3$OH, SiO, NH$_3$) close to high-mass protostars and young stellar objects (YSOs) can be detected and imaged in the radio regime. Masers are ideal probes of early-stage high-mass star formation as they arise in compact bright spots that can easily be tracked with VLBI to study gas kinematics with sub-milliarcsecond angular resolution \citep[e.g.][]{ torrelles, goddi5, sanna, goddi2, moscadelli}. They generally require velocity coherence along the line of sight and dense warm (or hot) gas for excitation. Different molecules enable studies of different regions in close proximity to high-mass compact objects \citep{elitzur, cragg, goddi, hollenbach}. 

The closest known example of a high-mass YSO is Source~I, an approximately equal-mass 20\,M$_\odot$ binary YSO \citep{goddi2}. It is deeply embedded in the Orion BN/KL nebula, at a distance of $418\pm6$\,pc \mbox{\citep{ment, kim}}. Exceptionally, Source~I excites SiO masers, which are usually only found in the circumstellar shells of evolved red giant stars. The Orion $^{28}$SiO masers were first discovered in 1973 \citep{snyd} and linked to Source~I by \cite{ment2}, making BN/KL the first of seven known high-mass star-forming regions to date where SiO masers have been observed \citep{hasegawa, zapata, ginsburg, higuchi, cho2}. These masers are particularly interesting because they require high temperatures and total gas densities -- 1000-3000\,K and $O(10^{9-10})$\,cm$^{-3}$ \citep{goddi}, and thus are excited at small radii, thereby providing a unique probe of gas dynamics and energetics \citep{green98, doel7, eisner, kim}.

SiO masers in Source~I have been previously observed with VLBI at 43\,GHz ($\lambda$7mm) and 86\,GHz ($\lambda$3mm). \cite{matt} imaged the emission from the $\lambda$7mm $^{28}$SiO $J=1-0$ rotational transition in the $v=1,2$ vibrational states based on a multi-epoch monitoring study in 2001--2002 and demonstrated that the masers arise at the surface of an (almost) edge-on disk and a wide-angle rotating outflow, at projected distances of 20--100\,AU from the central YSO. The first VLBI observations of the $\lambda$3mm ($v=1$, $J=2-1$) $^{28}$SiO maser emission were carried out with the Coordinated Millimeter VLBI Array but resulted in a single short-baseline (85\,km) detection, unsuitable to recover the maser flow structure with imaging \citep{doel}. Nevertheless, this early experiment enabled a comparison between the bulk of maser emission at $\lambda$7mm \citep{doel7} and Gaussian-fitted average positions of $\lambda$3mm emission, which suggested that the $\lambda$3mm SiO emission occurs further away from the YSO than the $\lambda$7mm emission. We have used the National Radio Astronomy Observatory's\footnote{The National Radio Astronomy Observatory is a facility of the National Science Foundation operated under cooperative agreement by Associated Universities, Inc.} (NRAO) Very Long Baseline Array (VLBA) to observe and image for the first time the $\lambda$3mm SiO emission associated with Source~I. The resulting images have a resolution comparable to the $\lambda$7mm images by \cite{matt}. 

The organization of the paper is as follows. After summarizing the observations and data reduction process (Section~\ref{sec:data}), we present our VLBA images and compare our results with the $\lambda$7mm images by \cite{matt} in Section~\ref{sec:results}.
In Section~\ref{sec:discussion}, we discuss a suitable model for the gas kinematics around Source~I (Section~\ref{sec:model}) as well as physical conditions in the SiO maser flow predicted by maser pumping models (Section~\ref{sec:pump}). A summary is given in Section~\ref{sec:summary}. 

\begin{figure*}
\begin{minipage}[t]{0.34\textwidth}
\includegraphics[width=\linewidth]{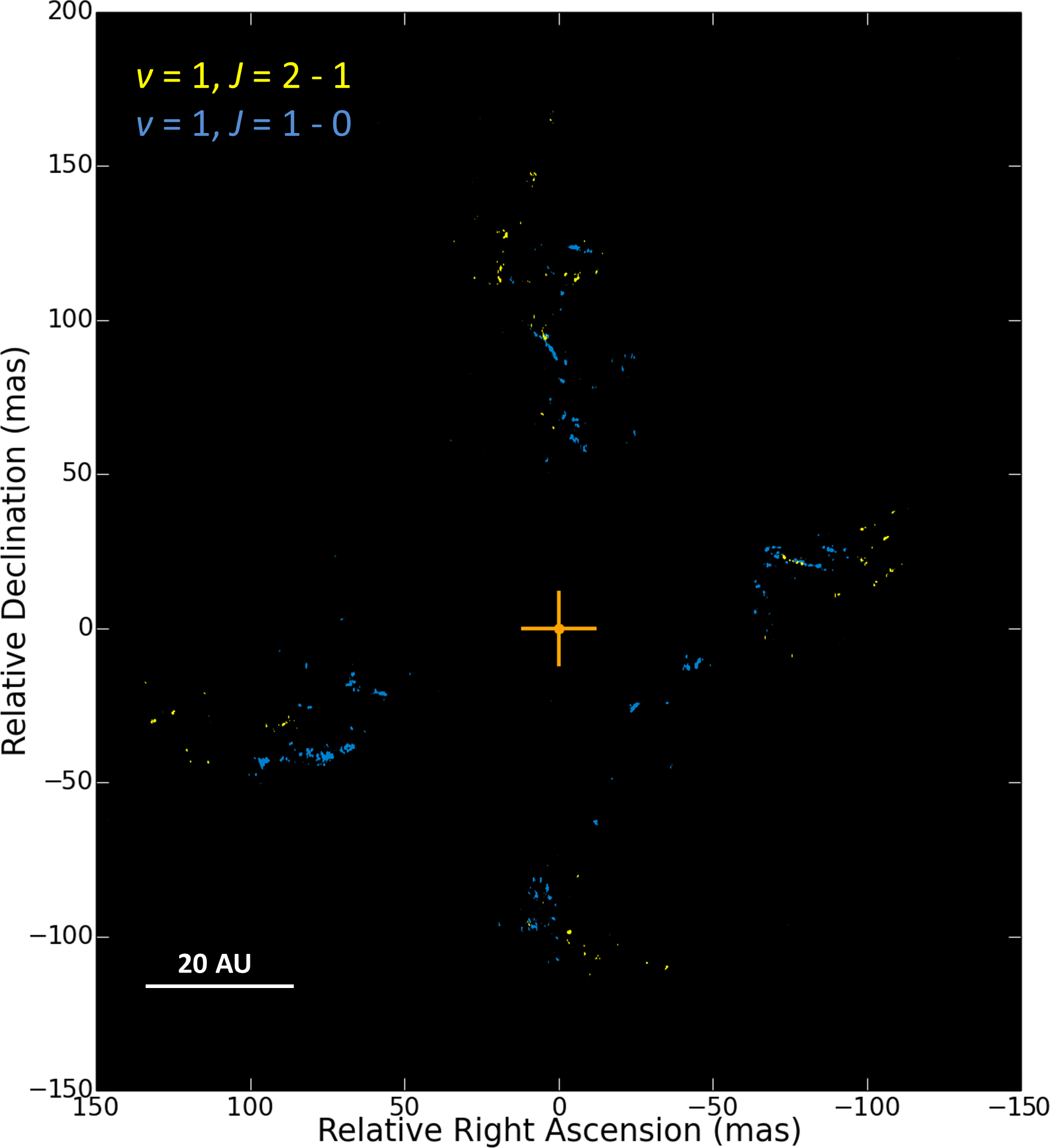}
\end{minipage}
\begin{minipage}[t]{0.33\textwidth}
\includegraphics[width=\linewidth]{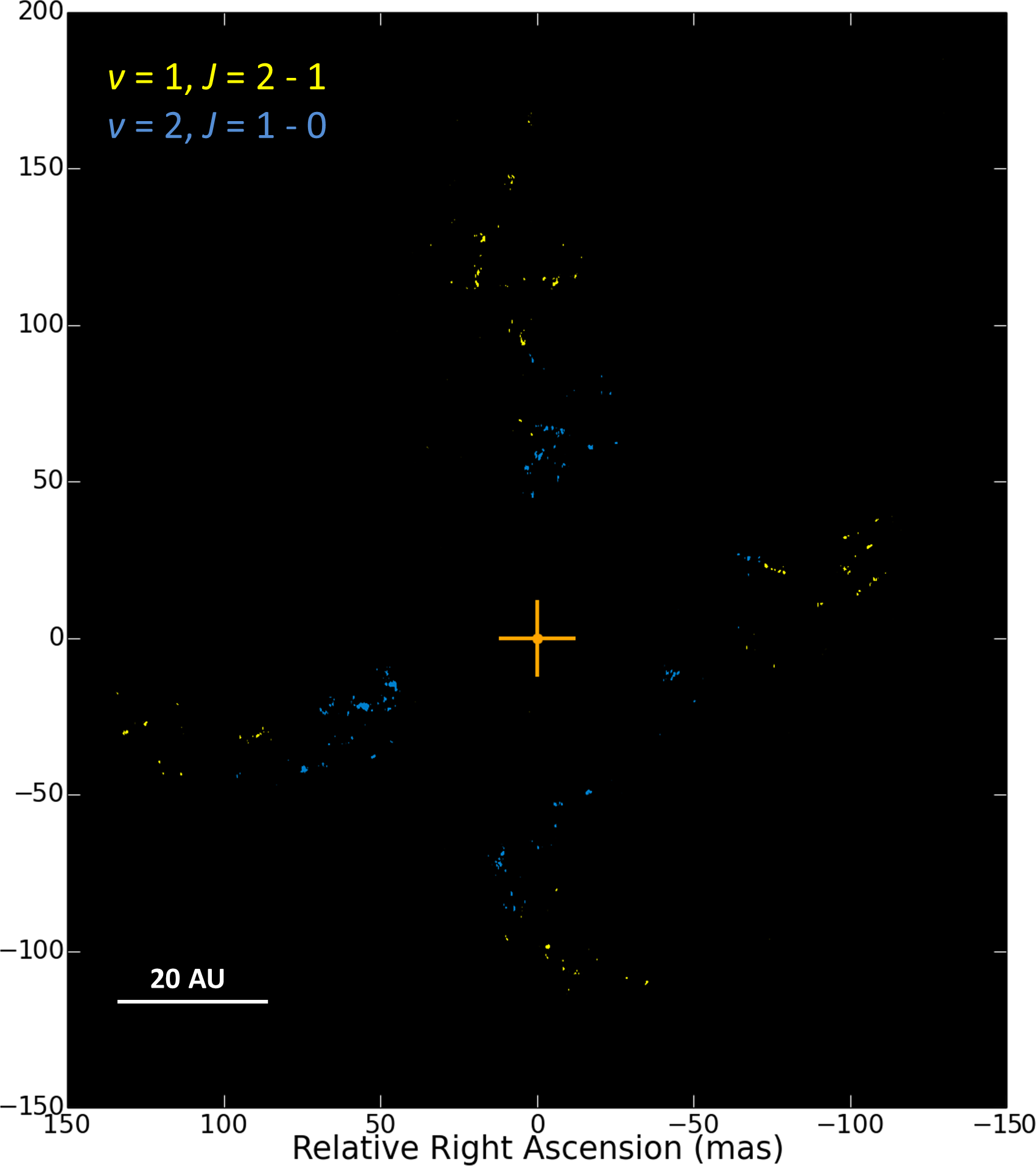}
\end{minipage}
\begin{minipage}[t]{0.33\textwidth}
\includegraphics[width=\linewidth]{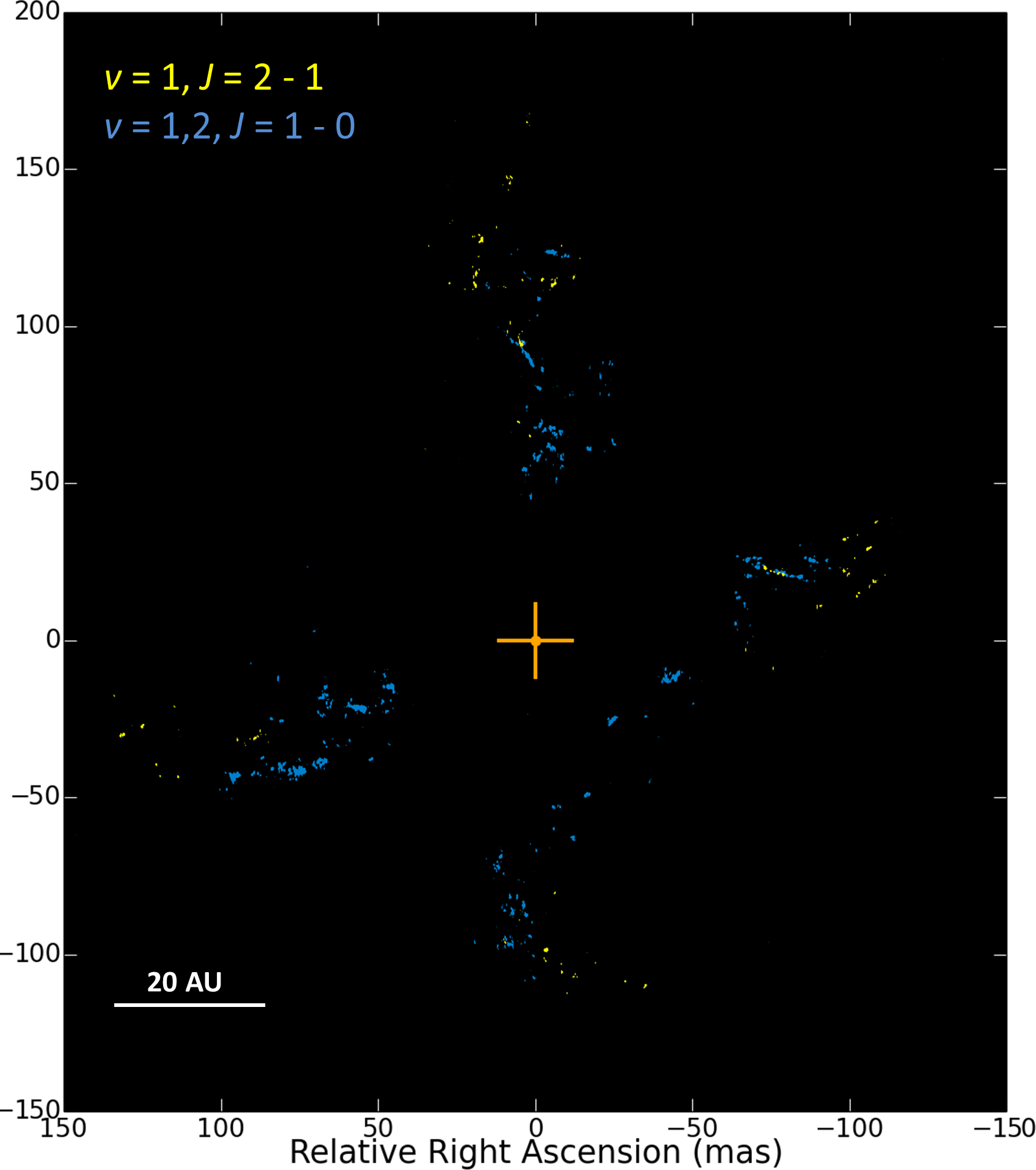}
\end{minipage}
\caption{Overlay of the velocity-integrated total-intensity images of $^{28}$SiO maser transitions at $\protect\lambda$3mm (in yellow, observed on January 24, 2011, integrated over the velocity range of $-$14.2 to $+$27.9\,km\,s$^{-1}$) and $\protect\lambda$7mm (in blue, observed on October 7, 2002, integrated over the velocity range of $-$48.1 to $+$62.3\,km\,s$^{-1}$): \textit{(left)} $\protect\lambda$3mm $v=1$, $J = 2-1$ and $\protect\lambda$7mm $v=1$, $J=1-0$; \textit{(center)} $\protect\lambda$3mm $v=1$, $J = 2-1$ and $\protect\lambda$7mm $v=2$, $J=1-0$; \textit{(right)} $\protect\lambda$3mm $v=1$, $J = 2-1$ and both $\protect\lambda$7mm $v=1$ and $v=2$, $J=1-0$. The cross indicates the estimated position for Source~I.}
\label{fig:mom0_overlay}
\end{figure*}

\section{Data reduction and imaging}\label{sec:data}
Observations of the $^{28}$SiO $v=1$, $J=2-1$ transition toward Source~I ($\alpha_{\mathrm{J}2000} = 05^\mathrm{h}35^\mathrm{m}14^\mathrm{s}.5149$, $\delta_{\mathrm{J}2000} = -05^{\circ}22'30''.582$) were made with all eight of the VLBA antennas equipped with 86\,GHz receivers on January 24, 2011 (project code BG205). The observations were carried out in a single polarization (RR) with a total bandwidth of 64\,MHz continuous over four intermediate frequency bands (IFs) to search for high-velocity components of the $^{28}$SiO $v=1$, $J=2-1$ transition ($v_\mathrm{rest} = 86.2434$\,GHz).\footnote{None were found. The velocity range of detected  $^{28}$SiO $v=1$, $J=2-1$  emission was from $-$12.3 to $+$25.4\,km\,s$^{-1}$ (in the third IF). See $\S3$.} The data were processed with the VLBA correlator, yielding 1024 spectral channels per IF with a spectral channel spacing of $\sim$0.054\,km\,s$^{-1}$.  The 8-hour track included four calibrator sources: 0528+134; 0458-020; 0539-037; and 0605-085. The total integration time on Source~I was of 5.26\,hours.

Data reduction utilized the NRAO Astronomical Imaging Processing System (AIPS). Visibilities with spurious amplitudes or phases were flagged. Parallactic angle corrections were applied to phases to correct for feed rotation on the sky at each antenna during the observations, and digital sampler bias corrections were applied to  antenna-based amplitude offsets introduced by variations in  transition voltages. 

Atmospheric delay and instrumental delay and phase offsets were estimated for each IF of each antenna from a fringe fit on a five-minute calibrator scan (0528+134) where all antennas were present, using the task FRING.
Due to the coherence time at 86\,GHz being relatively short in comparison to the scan length, we looked for solutions every 30 seconds. Channel-averaging the data down to 128 channels also enabled a higher number of solutions found during the fringe search. Fringe solutions were found in all baselines except those to the Mauna Kea (MK) VLBA station, which were consequently flagged. Delay residuals and rates corrections were done using fringe fits on all available calibrator scans. Delay and rate solutions were then copied from the frequency-averaged data and applied to the original dataset. 

The amplitude component of the bandpass calibration was estimated using the total-power spectra of the calibrator 0528+134 for each antenna, with the task BPASS, and applied to cross-power spectra for Source~I. The phase portion of the bandpass calibration was estimated by fitting a complex polynomial to the cross-power spectra of the same calibrator with the task CPASS. Subsequently, preliminary amplitude calibration was carried out using standard NRAO gain curves for the antennas and system temperatures measured during the observations. Cross-power spectra for Source~I were also Doppler-shifted to compensate for the Earth's motion, using the task CVEL, and establishing a velocity scale in the Local Standard of Rest (LSR). 

Next, residual fringe-rates were determined using a bright maser feature in an individual spectral channel as a reference. The solutions obtained in the reference channel were then applied to all spectral channels. Self-calibration was performed on the reference channel of the chosen maser feature first in phase only and then in both amplitude and phase. This was carried out by iteratively self-calibrating the single-channel data, using the task CALIB, and deconvolving interactively using CLEAN in the task IMAGR. 
The solutions were then applied in two steps, first phase-only and then both amplitude and phase, to the full spectral-line dataset. 

An automated imaging and deconvolution of the dataset was executed using the task IMAGR, producing a line cube covering the whole velocity range where a signal was visible in the total-power spectra (from $-$14.2 to 27.9\,km\,s$^{-1}$). In order to reduce artifacts associated with strong side-lobes (as expected for an equatorial source with strong emission), the preliminary image cube was then interactively cleaned by carefully placing CLEAN boxes channel by channel. The resulting channel maps were used to replace individual channels in the original preliminary cube using the task MCUBE. All channels were cleaned to a level of 60\,mJy\,beam$^{-1}$. Imaging was done using a pixel size of 0.05\,mas and an image size of 8192 $\times$ 8192 pixels, comparable to the $\lambda$7mm images by \cite{matt}. The restoring beam used had a size of 0.55 $\times$ 0.19\,mas with a position angle of 0.0$^{\circ}$ (from north in an easterly direction). The thermal noise rms in the channels with no line emission ranged from 55 to 60\,mJy\,beam$^{-1}$ and the rms in the channels with strong line emission ranged from 80 to 90\,mJy\,beam$^{-1}$. Regional (over specific ranges of RA and Dec) and global (over the whole cube) spectra were extracted from the image cube using the task ISPEC. 
Individual components of the emission structure were extracted using the task SAD, by searching for emission and fitting two-dimensional ellipsoidal Gaussian models to each channel map of the SiO maser source above a detection threshold varying across channels, depending on the ratio of the strongest feature to the thermal noise rms in that channel (starting at 3$\sigma$ for channels where the ratio is below 10). This was adopted to control for the limited dynamic range of the images. 
This method identified 120 distinct maser features. The term `feature' here indicates a collection of spectrally and spatially contiguous maser spots. A maser feature is considered real if it is detected in at least five contiguous channels (for a typical SiO line width of $\sim$0.3\,km\,s$^{-1}$), with a position shift of the intensity peak from channel to channel smaller than the FWHM size. Positions of the individual maser features were obtained by taking the intensity-weighted mean of the locations of the corresponding collection of Gaussian fits and radial velocities were determined by taking the intensity-weighted mean velocity \citep[for more details on this method see][]{goddi6}.

After imaging, the final cube was exported from AIPS and imported into the Common Astronomy Software Applications package (CASA) for moment analysis. The final zeroth (velocity-integrated total intensity) and first (velocity field) moment maps were generated using the CASA task \textit{immoments}, removing channels with no emission and excluding pixels below the thermal noise rms of channels with no emission. The zeroth and first moments were generated with pixel flux thresholds of 0.1\,Jy and 1.2\,Jy respectively. 

\section{Results}\label{sec:results}
We obtained images of the $\lambda$3mm $v=1$, $J=2-1$ $^{28}$SiO maser emission observed with the VLBA on January 24, 2011. The following sections describe the resulting images and comparisons with the $\lambda$7mm $v=1$ and $v=2$, $J=1-0$ $^{28}$SiO maser emission observed with the VLBA in the monitoring campaign from 2001--2002 \citep{matt}. We compared the maser emission structures using total intensity images (Section~\ref{sec:structure}) and velocity fields (Section~\ref{sec:field}). We additionally characterized individual maser features from the Gaussian fits, measured velocity gradients and estimated spatial offsets between different transitions (Section~\ref{sec:gauss}). Furthermore, we used the resulting images to extract regional and global VLBI spectral profiles of the $\lambda$3mm emission and compare them with the two transitions of the $\lambda$7mm emission as well as with the total-power spectral profiles (Section~\ref{sec:profiles}). 

\subsection{The SiO maser emission structure}\label{sec:structure}
The velocity-integrated total-intensity map of the $^{28}$SiO $v=1$, $J=2-1$ transition emission at $\lambda$3mm was registered with velocity-integrated images of two $\lambda$7mm transitions ($v=1$ $\&$ $v=2$, $J=1-0$) and centered on Source~I (shown in Fig.~\ref{fig:mom0_overlay}). The putative position of Source~I (center cross) in the images of individual transitions was estimated using the intersection of linear fits of the north-south and east-west maser regions, using maser emission intensity as weights. For a one-to-one comparison, a single epoch from the 2001--2002 monitoring campaign at $\lambda$7mm was selected, observed on October 7, 2002 \citep{matt}.

Based on the overlays shown in Fig.~\ref{fig:mom0_overlay}, we can assess that the $^{28}$SiO $\lambda$3mm emission has an X-shaped distribution, consistent with that known for $\lambda$7mm emission. In particular, the $\lambda$3mm masers lie in four distinct regions, or arms, extending to the north, west, south and east, and centered on Source~I. Interestingly, although some overlap is present between the $\lambda$3mm and $\lambda$7mm emission in the four arms (Fig.~\ref{fig:mom0_overlay}, right panel), the $\lambda$3mm emission appears to systematically prefer larger radii further away from Source~I, as first noticed by \cite{doel}. No $\lambda$3mm emission is seen in the inner regions, at radii less than 25\,AU from the position of Source~I, nor in the bridge observed in the $\lambda$7mm image. Here the bridge refers to the emission connecting the south and west arms at $\lambda$7mm, and in the model presented in Fig.~2 of \cite{matt} it traces the limb-brightened edge of a rotating disk. 

\begin{figure}
\includegraphics[width=\linewidth]{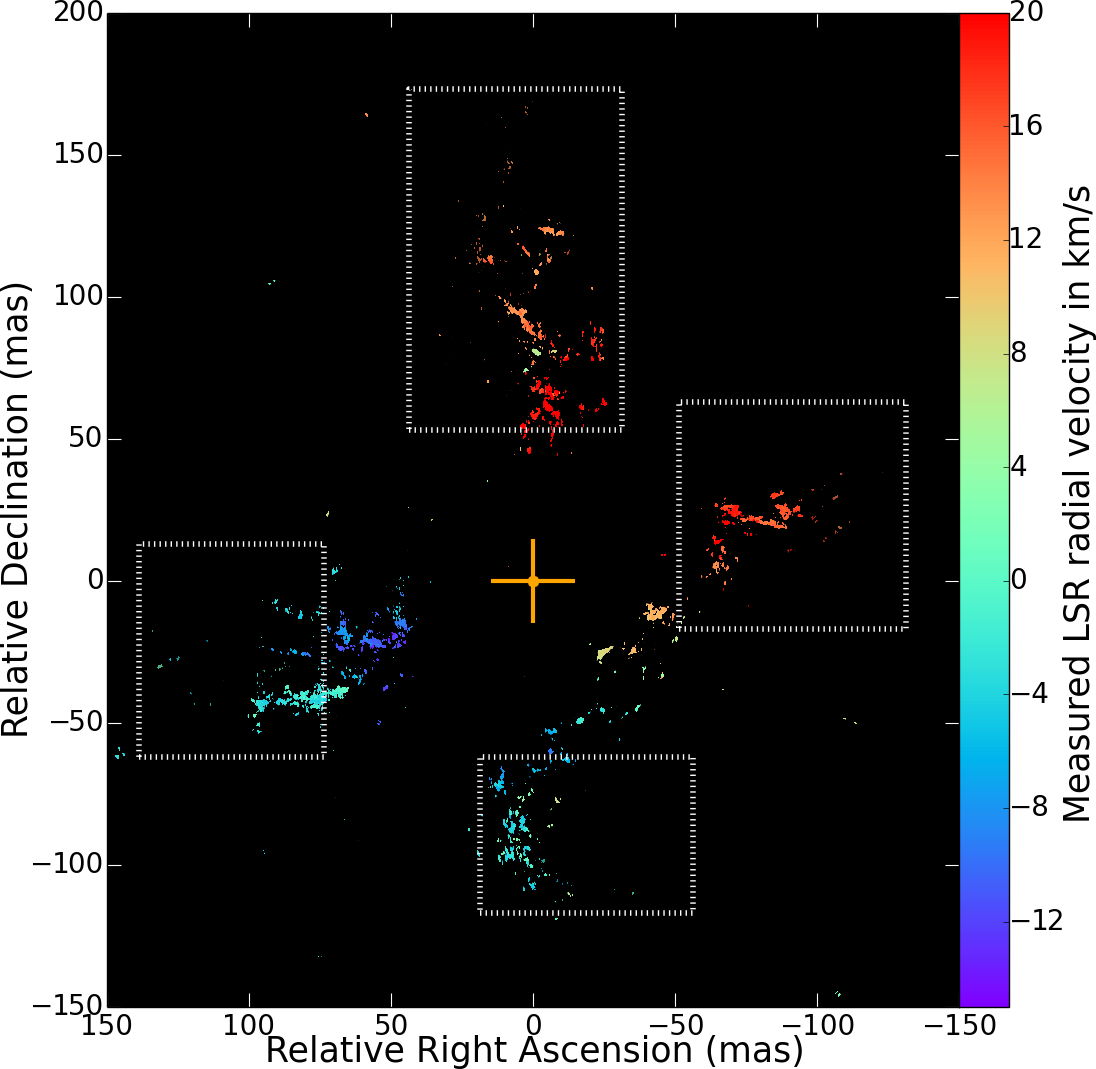}
\caption{Overlays of the velocity field images of the $\lambda$3mm $^{28}$SiO $v=1$, $J=2-1$ emission of January 24, 2011 and both the $\lambda$7mm $^{28}$SiO $v=1$, $J=1-0$ and $v=2$, $J=1-0$ maser emission of October 7, 2002. The same color scale is used for both wavelengths, and the velocity fields seem continuous and consistent with each other. The cross indicates the estimated position for Source~I. The white boxes delimit the extent of the $\lambda$3mm emission.}
\label{fig:7mm_MOM1}
\end{figure}

Regarding individual transitions, the overlay of the $\lambda$3mm velocity-integrated total-intensity image with that of the $\lambda$7mm $v=2$ (Fig.~\ref{fig:mom0_overlay}, center panel) shows significantly less overlap than the overlay with the $\lambda$7mm $v=1$ image (Fig.~\ref{fig:mom0_overlay}, left panel). This suggests that there should be regions where the $J=2-1$ transition at $\lambda$3mm occurs in similar gas volumes to the $J=1-0$ transition at $\lambda$7mm of the same vibrational state, but in different gas volumes to that of a higher vibrational state ($v=2$). However, due to the overall spatial offset observed between the bulk of the $\lambda$3mm and the $\lambda$7mm masers, even rotational transitions in the same vibrational state likely require different ranges of physical conditions for optimal excitation. Based on this finding, it is reasonable to assume that regions with a spatial overlap between the $v=1$ transitions at $\lambda$3mm and $\lambda$7mm likely have a set of density and temperature ranges where physical conditions are favorable for both transitions, while regions with no spatial overlap exhibit conditions favorable to only one of the two. In particular, we expect the $v=1$, $J=2-1$ and the $v=2$, $J=1-0$ transitions to have distinct physical conditions since no overlap is found. 

A caveat is that the high-resolution $\lambda$7mm observations were carried out ten years before these $\lambda$3mm observations. We thus cannot exclude that variability in the source structure over 10 years may affect a cross-comparison between the two transitions. Nevertheless, the good correspondence between the observed spatial and velocity structures (see also Section~\ref{sec:field}) suggests that time variability does not likely affect the bulk structure of the emission at the two frequencies. Therefore, we are confident that the overlays presented here are a good approximation of the spatial correspondence between the $\lambda$3mm and $\lambda$7mm maser emission. 

\begin{figure}
\includegraphics[width=\linewidth]{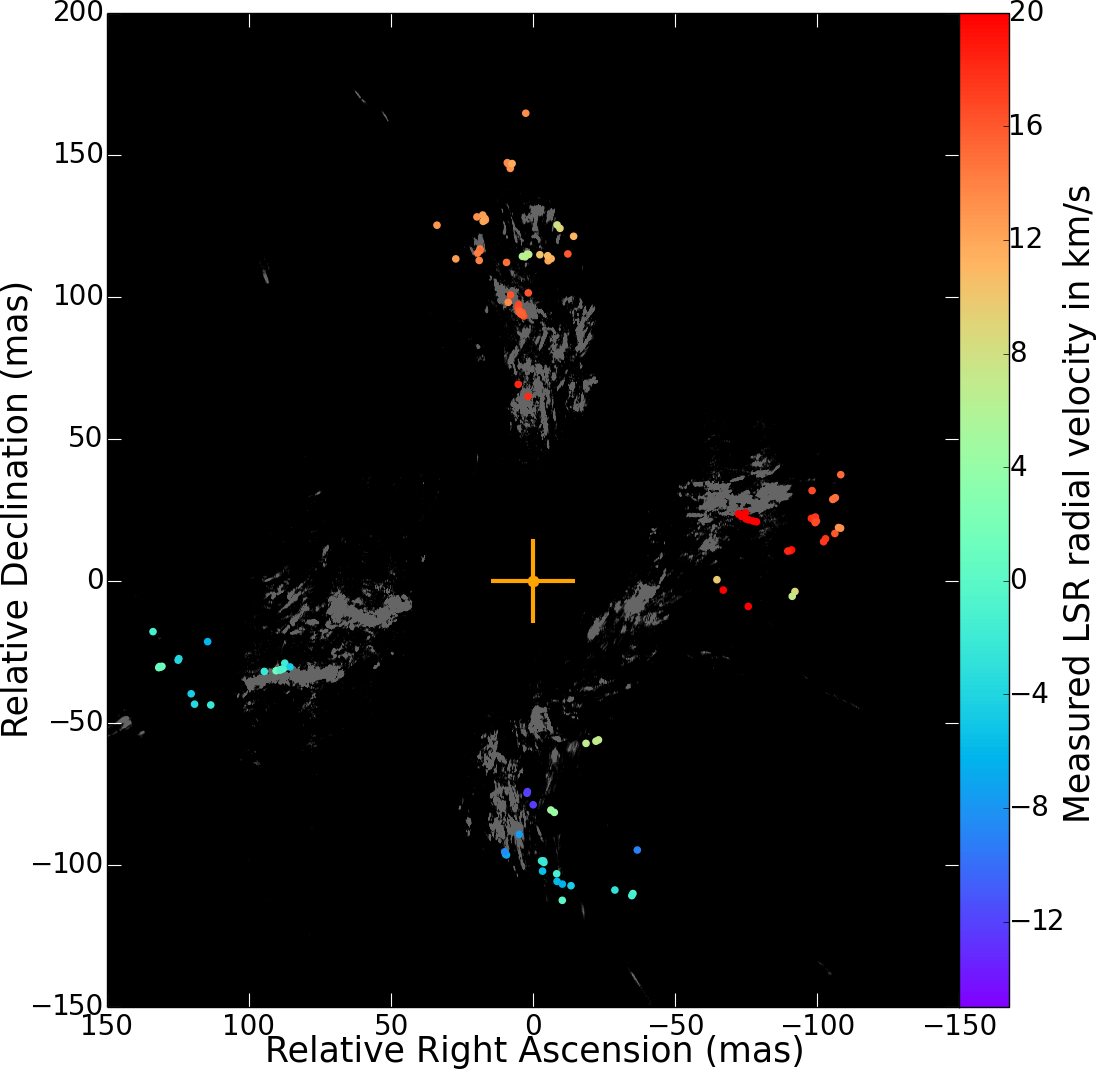}
\caption{Gaussian-fitted positions and velocities of individual maser features of the $^{28}$SiO $v=1$, $J=2-1$ emission, overlaid on the velocity-integrated intensity map of the $\protect\lambda$7mm emission \protect\citep[all 19 epochs of data spanning 22 months, in grey-scale;][]{matt}. The cross indicates the estimated position  for Source~I. Weak $\lambda$3mm emission not visible in Fig.~\ref{fig:mom0_overlay} was recovered in the north and south arms and west of the base of the south arm.}
\label{fig:sad}
\end{figure}

\begin{table*}[!ht]
\caption{Measured centroid distances of the $^{28}$SiO $\lambda$3mm maser emission (obtained from the Gaussian features and the velocity-integrated intensity map) and the $\lambda$7mm $v=1,2$ maser emission (obtained from the velocity-integrated intensity maps) from Source~I, and offsets between the $\lambda$3mm and $\lambda$7mm maser emission of each arm from the velocity-integrated intensity maps.}
\label{tab:offsets}
\centering
\small
\begin{tabularx}{\linewidth}{Xcccccc}
\hline \hline
\noalign{\smallskip}
Region &  $R_\mathrm{Gauss}^{(a)}$ [AU] & $R_\mathrm{map}^{(b)}$ [AU]&  $R_\mathrm{map}$ [AU] &  $R_\mathrm{map}$ [AU] & $\Delta R_\mathrm{map}^{(c)}$ for $\lambda$3mm $v=1$  & $\Delta R_\mathrm{map}$ for $\lambda$3mm $v=1$  \\
  & ($\lambda$3mm $v=1$)  &  ($\lambda$3mm $v=1$) &  ($\lambda$7mm $v=1$)  &  ($\lambda$7mm $v=2$) & and $\lambda$7mm $v=1$ [AU] &  and $\lambda$7mm $v=2$ [AU]  \\
\noalign{\smallskip}
\hline
\noalign{\smallskip}
North & 48 $\pm$ 8  & 52 $\pm$ 2 & 44 $\pm$ 8 & 35 $\pm$ 7  & 8  & 17  \\

West & 39 $\pm$ 6 & 40 $\pm$ 9 & 34 $\pm$ 6 & 34 $\pm$ 6  & 6  & 6  \\

South & 41 $\pm$ 5 & 39 $\pm$ 2 & 36 $\pm$ 5 & 32 $\pm$ 5 & 3 & 7  \\

East & 48 $\pm$ 8 & 47 $\pm$ 9 &  35 $\pm$ 7 & 34 $\pm$ 7  & 12  & 13 \\ 
\noalign{\smallskip}
\hline
\end{tabularx}
\tablefoot{$^{(a)}$ Centroid distances per arm obtained from the Gaussian features; $^{(b)}$ centroid distances per arm obtained from the velocity-integrated intensity maps; $^{(c)}$ relative offsets of emission centroids obtained from the velocity-integrated intensity maps.}
\end{table*}

\subsection{The SiO velocity field}\label{sec:field}
We registered the velocity field map of the $\lambda$3mm $v=1$ emission with a single epoch $\lambda$7mm velocity field ($v=1$ $\&$ $v=2$; Fig.~\ref{fig:7mm_MOM1}). 
This map clearly shows that the east and south arms are blue-shifted while the north and west arms are red-shifted around a systemic velocity for Source~I estimated at $v_\mathrm{sys} = 5.5$\,km\,s$^{-1}$. 
Moreover, the maser emission at $\lambda$3mm and $\lambda$7mm cannot be easily differentiated from each other as they show considerable overlap in position and velocity spread, and show a continuous velocity gradient in all four arms. Nonetheless, the $\lambda$3mm emission (delimited by the dashed boxes in Fig.~\ref{fig:7mm_MOM1}) has a smaller velocity range ($-$12.3\,km\,s$^{-1}$ to $+$25.4\,km\,s$^{-1}$) than the $\lambda$7mm emission ($-$14.1\,km\,s$^{-1}$ to $+$26.5\,km\,s$^{-1}$) and lies, on average, at larger projected distances from Source~I (as first noticed in \S~\ref{sec:structure}). This is consistent with a Keplerian rotation velocity profile, where faster rotation occurs closer to the YSO. This strongly supports the hypothesis that both the $\lambda$3mm and the $\lambda$7mm SiO maser emission must come from the same wide-angle outflow, undergoing the same Keplerian rotation around a northeast-southwest rotation axis (see full discussion in \S~\ref{sec:model}). 

\begin{figure}
\centering
\includegraphics[width=\linewidth]{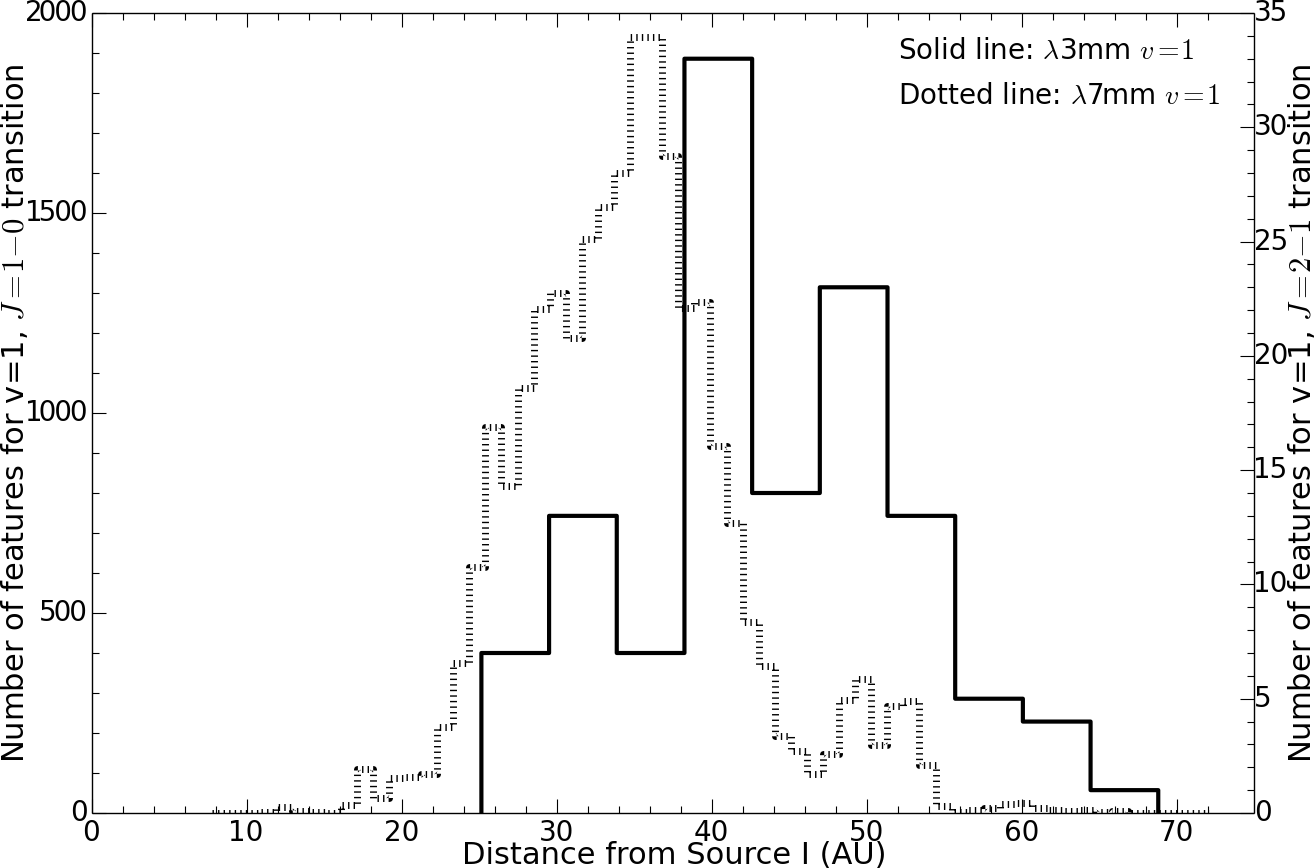}
\caption{Histograms showing the total number of individual  maser features observed in the $\protect\lambda$3mm $v=1$, $J=2-1$ (January 24, 2011) and $\protect\lambda$7mm $v=1$, $J=1-0$  \protect\citep[all 19 epochs;][]{matt} emission and their positions relative to Source~I. The histogram bin width is 1\,AU for $\lambda$7mm, and 5\,AU for $\lambda$3mm due to the low number of features.}\label{fig:gauss}
\end{figure}

\subsection{The individual maser features at $\lambda$3mm}\label{sec:gauss}
We report positions and LSR velocities of the Gaussian-fitted individual maser features of the $\lambda$3mm $v=1$, $J=2-1$ transition in Fig.~\ref{fig:sad}, where they are overlaid on the velocity-integrated intensity emission at $\lambda$7mm \citep{matt}. The $\lambda$7mm map comprises a combination of 19 epochs of $v=1$ and $v=2$ data spanning 22 months. The Gaussian-fitting technique enabled the recovery of low-intensity features in channels with strong emission and in channels where no emission was found in the moment maps. Weaker features were recovered in the north and south arms and most notably three individual components were extracted in the region between the south and west arms, close to the base of the south arm. These features do not seem to correspond to the broad structure of the south arm and have radial velocities near the systemic velocity. This could possibly be a component of a $\lambda$3mm bridge, similar to the bright bridge present at $\lambda$7mm (e.g. Fig.~\ref{fig:7mm_MOM1}). These features do not however possess a $\lambda$7mm counterpart at a similar velocity nor in the same region. It is thus likely that these new features are sampling a different volume of gas than the bridge emission at $\lambda$7mm.\\
We measured distances of the extracted individual features from the central object Source~I at $\lambda$3mm and we compared the resulting spatial distribution to the $\lambda$7mm $v=1$ distribution by \cite{matt}, shown in Fig.~\ref{fig:gauss}. There is significant overlap between the two distributions, suggesting that the two transitions at $v=1$ may share common radii of excitation from the source. However, the peak of the $\lambda$3mm distribution is offset from the $\lambda$7mm peak by 5--15\,AU, a result consistent with the position offsets estimated by \cite{doel}. 

We additionally measured centroid distances from Source~I in each arm both for individual features  and from the velocity-integrated total-intensity map, using maser intensities as weights. These centroid distances are on average 50\,AU for the north and east arms and 40\,AU for the south and west arms (Table~\ref{tab:offsets}), which correspond to the peaks of the $\lambda$3mm maser feature distribution (Fig.~\ref{fig:gauss}). The intensity map technique was also used to measure centroid distances of each arm for the single-epoch $\lambda$7mm $v=1$ and $v=2$ velocity-integrated intensity maps. Using these estimates, we  obtained the spatial offsets between the $\lambda$3mm and $\lambda$7mm SiO maser emission, presented in Table~\ref{tab:offsets}. 
This provides a quantitative confirmation of the statement (made in Section~\ref{sec:structure}) that the bulk of the $\lambda$3mm $v=1$ is closer overall to the $\lambda$7mm $v=1$ emission ($\sim$7\,AU offset) than to the $v=2$ emission ($\sim$11\,AU offset). We also note that the offsets between the $v=1$ $\lambda$3mm and $\lambda$7mm emission centroids found with the velocity-integrated intensity map method are consistent with the peak offset determined from the feature distribution histograms in Fig.~\ref{fig:gauss}. 

\begin{figure*}[ht]
 \centering
\includegraphics[width=0.81\textwidth]{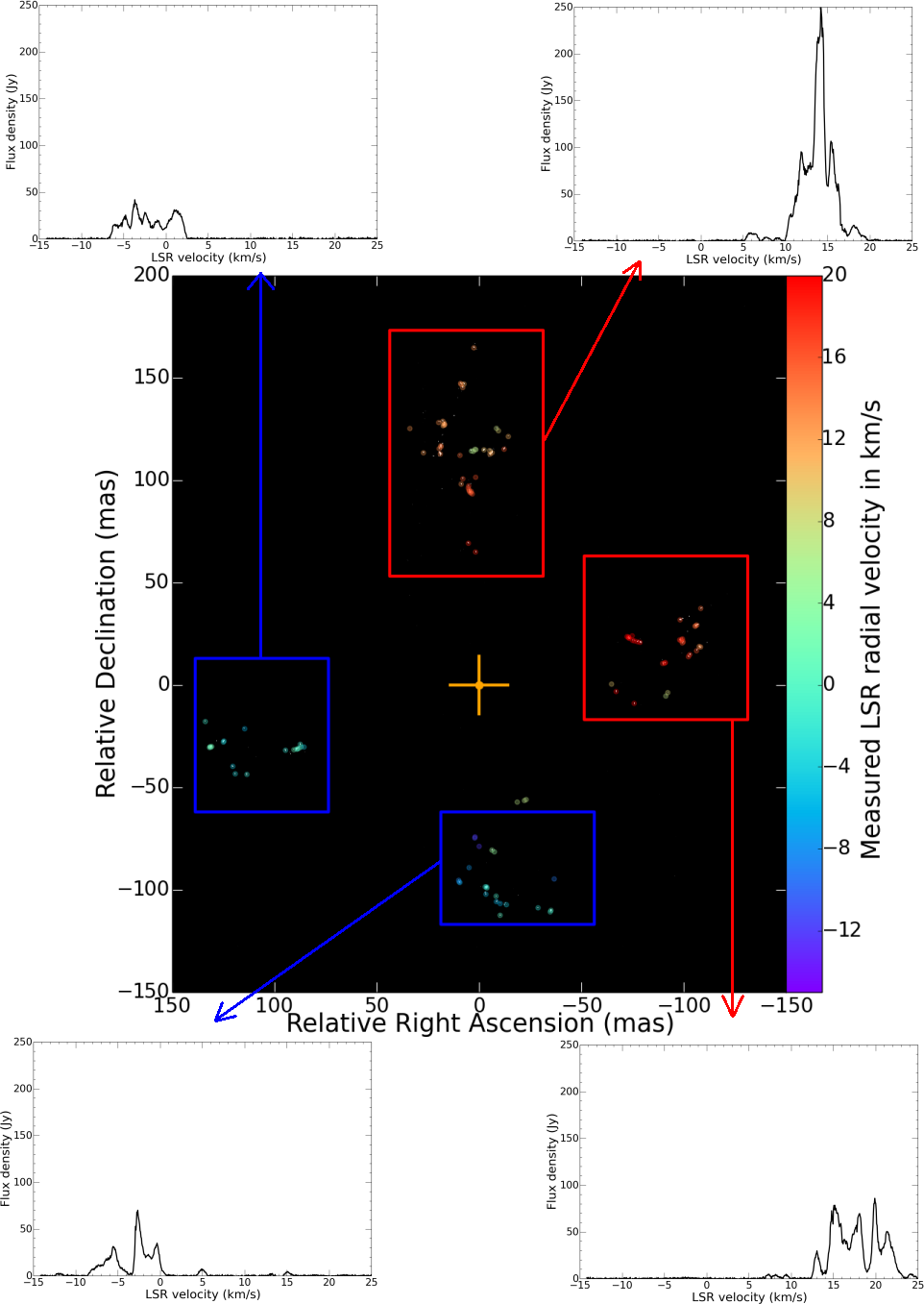}
\caption{Positions and velocities of individual maser features of the $^{28}$SiO $v=1$, $J=2-1$ maser emission around Source~I observed on January 24, 2011. Spectra corresponding to delimited regions on the map are included, per individual arm of the emission structure.}\label{fig:arms}
\end{figure*}

Finally, we used the Gaussian-fitted individual maser features to estimate velocity gradients across each arm. Both red-shifted arms (north and west) and the blue-shifted south arm have measurable velocity gradients, consistent with a deceleration as a function of distance from the source. No velocity gradient was observed in the east arm. In particular, the red-shifted north and west arms have individual velocity gradients of 0.17$\pm$0.05\,km\,s$^{-1}$\,AU$^{-1}$ and 0.4$\pm$0.1\,km\,s$^{-1}$\,AU$^{-1}$ respectively and the blue-shifted south arm has a velocity gradient of 0.3$\pm$0.2\,km\,s$^{-1}$\,AU$^{-1}$. These correspond to an overall velocity gradient of 0.3$\pm$0.1\,km\,s$^{-1}$\,AU$^{-1}$, consistent with the single-arm gradient estimated by \cite{doel} and the overall gradient at $\lambda$7mm measured by \cite{matt}. A single overall gradient for both the $\lambda$3mm and $\lambda$7mm SiO maser emission points towards a global property of the outflow, notably a differential rotation with a Keplerian velocity profile, rather than a deceleration due to a localized phenomenon.

\subsection{The SiO spectral profiles}\label{sec:profiles}
\subsubsection{The regional spectral profiles at $\lambda$3mm}\label{sec:vel}

We extracted subsets of spectra from the line cube in different spatial regions, in order to assess the velocity and intensity contributions of each distinct arm to the total $^{28}$SiO $v=1$, $J=2-1$ emission spectrum around Source~I, as illustrated in Figure \ref{fig:arms}. Individual spectra of emission for the four distinct regions show similar velocity ranges between the two blue-shifted arms (south and east) and the two red-shifted arms (north and west). The velocity ranges of emission in each arm are reported in Table~\ref{tab:arm_vel}, in both the LSR frame and in the reference frame of Source~I ($v_\mathrm{sys}= 5.5$\,km\,s$^{-1}$).

Furthermore, each regional spectrum covers a similar velocity range to previous observations at $\lambda$3mm \citep{doel}, suggesting that there is no obvious long-term evolution of the velocity distribution of the $\lambda$3mm maser emission over the last 10 years.

\begin{table}[!h]
\caption{Velocity ranges of the individual spectra for the four arms of the $v=1$, $J=2-1$ emission in both LSR and in the reference frame of Source~I, centered at a systemic velocity of $v_\mathrm{sys}=5.5$\,km\,s$^{-1}$.}\label{tab:arm_vel}
\centering
\small 
\begin{tabularx}{\columnwidth}{lcc}
\hline
\hline 
\noalign{\smallskip}
Region & LSR velocity range & Source~I velocity range\\
 & [km\,s$^{-1}$] & [km\,s$^{-1}$]\\
\noalign{\smallskip}
\hline
\noalign{\smallskip}
North arm  & $+$6.2 to $+$18.4 & $+$0.7 to $+$12.9 \\

West arm & $+$7.2 to $+$25.4 & $+$1.7 to $+$19.9 \\

South arm  & $-$12.3 to $+$4.7  & $-$17.8 to $-$0.8 \\
East arm & $-$6.1 to $+$2.4  & $-$11.6 to $-$3.1 \\ 
\noalign{\smallskip}
\hline
\end{tabularx}
\end{table}

\begin{figure}
\centering
\includegraphics[width=\linewidth]{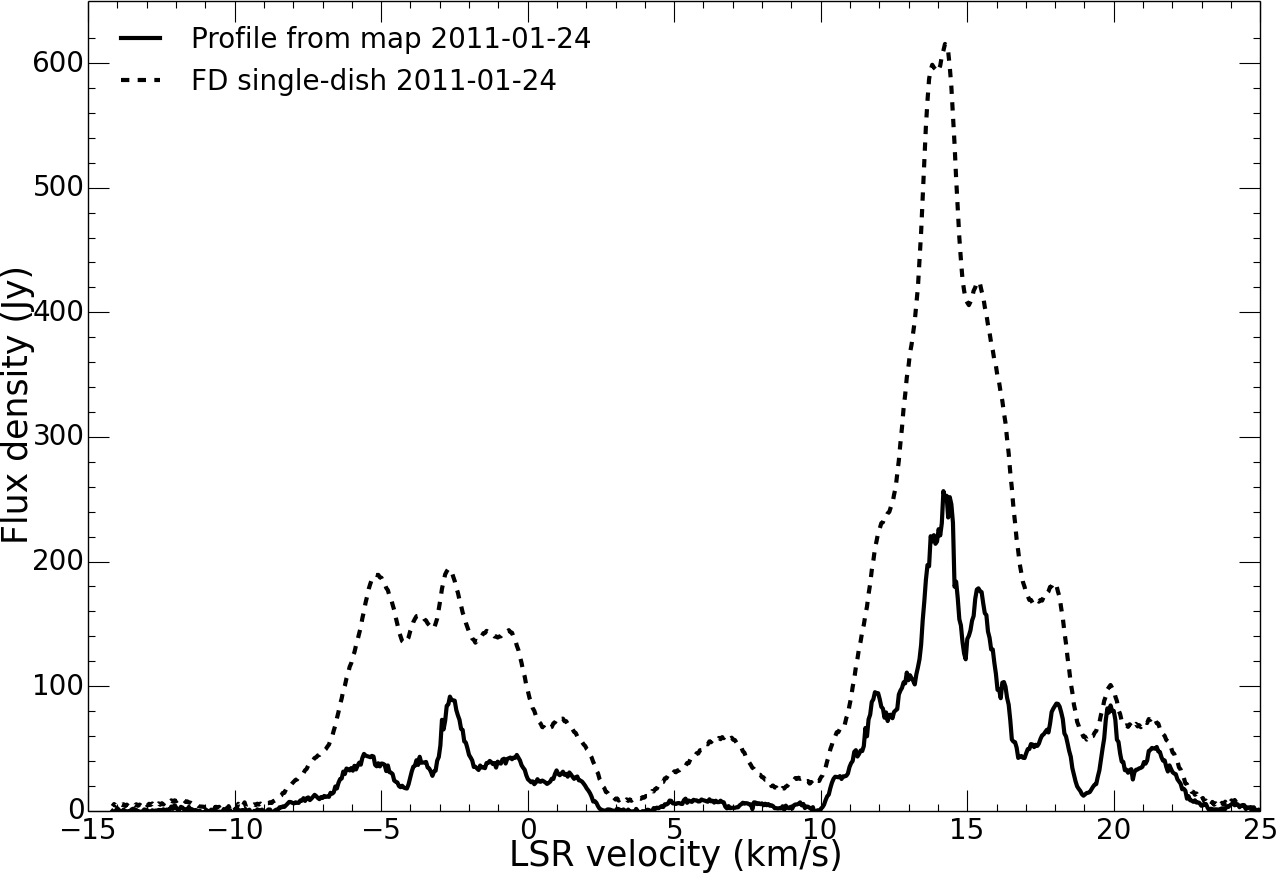}
\caption{Comparison of the $\lambda$3mm $v=1$, $J=2-1$ total-power spectral profile obtained from the VLBA station Fort Davis with the spectral profile obtained from the line cube shows noticeably similar features.}
\label{fig:FD}
\end{figure}

\subsubsection{The global integrated spectral profile at $\lambda$3mm}\label{sec:flux}
We compared the spectral profiles obtained from the total-power spectrum of the VLBA station Fort Davis and extracted from the line cube of the $\lambda$3mm $v=1$, $J=2-1$ maser emission (Fig.~\ref{fig:FD}). The total flux density recovery with VLBI compared to the total-power spectrum amounts to 33$\%$ of the full emission. In fact, data calibration resulted in the loss of two of the eight VLBA antennas observing at $\lambda$3mm (see Section~\ref{sec:data}). Additionally, the shortest baseline in the VLBA is roughly 250\,km, which likely prevents the recovery of more diffuse emission structures (the $\lambda$7mm monitoring experiments included a VLA antenna, providing a short baseline to Pie Town of $\sim$50\,km). This leads to the conclusion that a significant part of the missing flux density must come from extended or diffuse emission that is not sampled by the VLBA.

Despite the difference in flux level, the total-power and the VLBI spectra share similar peak locations and general profiles. Interestingly, there is a significant amount of flux density missing in the VLBI image-extracted spectrum near the systemic velocity, $v_\mathrm{sys}=5.5$\,km\,s$^{-1}$, possibly corresponding to the bridge regions between the north and east arms and between the south and west arms, where masers near the systemic velocity occur at $\lambda$7mm \citep{matt}. In total-power, the intensity of the emission near the systemic velocity ranges from 20\,Jy to 60\,Jy, which would correspond to fluxes of 6-20\,Jy in VLBI, assuming 33$\%$ recovery. If the systemic velocity emission consisted of bright compact spots of a few Jansky each, we  should still be able to detect them (1\,Jy corresponds to roughly 10$\sigma$ in our VLBI maps). Since the spectral peak near the systemic velocity is almost non-existent in the VLBI profile, this suggests that emission near $v_\mathrm{sys}$ may be composed of more diffuse or extended structures with respect to the blue- and red-shifted arms (larger than 3\,mas, corresponding to the resolution on the shortest Pie Town--Los Alamos baseline). 

\begin{figure}
\centering
\includegraphics[width=\linewidth]{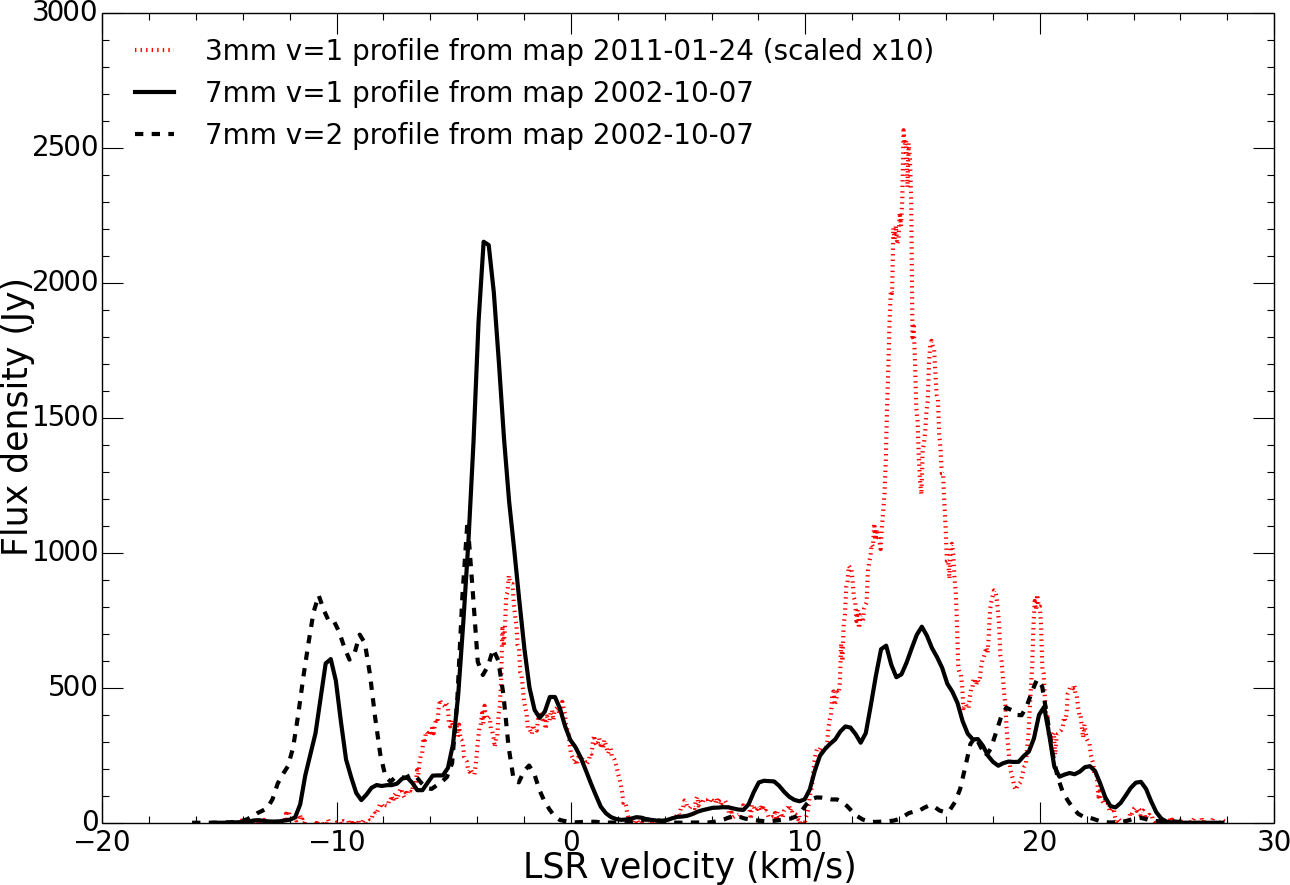}
\caption{Comparison of the $\lambda$3mm $v=1$, $J=2-1$ spectral profile from the line cube on January 24, 2011 (scaled by an order of magnitude) with the $\lambda$7mm $J=1-0$, $v=1,2$ spectral profiles from the line cubes on October 07, 2002.}
\label{fig:ISPEC}
\end{figure} 

\subsubsection{Comparisons of spectral profiles at $\lambda$3mm and $\lambda$7mm}\label{sec:comp}
We compared the total VLBI spectral profiles from the line cube for the $v=1$ observations at $\lambda$3mm to the single-epoch $v=1,2$ $\lambda$7mm observations (Fig.~\ref{fig:ISPEC}). The $\lambda$3mm spectrum is scaled by a factor of ten for a closer comparison to the much stronger emission from the $\lambda$7mm transitions. The $\lambda$3mm profile shows the characteristic double-peaked spectrum seen in both $J=1-0$ transitions at $\lambda$7mm. However, it is slightly narrower than the spectra at $\lambda$7mm: the $\lambda$3mm maser emission does not attain the higher blue-shifted and red-shifted velocities reached by masers at $\lambda$7mm. The total $\lambda$3mm velocity range was found to be from $-$12.3\,km\,s$^{-1}$ to $+$25.4\,km\,s$^{-1}$, which is overall lower than the velocity ranges for the $\lambda$7mm $v=1$ ($-$14.1\,km\,s$^{-1}$ to $+$26.5\,km\,s$^{-1}$) and $v=2$ ($-$15.3\,km\,s$^{-1}$ to $+$25.8\,km\,s$^{-1}$). Since the $\lambda$3mm masers are detected on average further from Source~I than the $\lambda$7mm emission, the narrower spectrum is consistent with Keplerian rotation of the gas, as already noticed.

A caveat is in place: the maser emission structure is variable on a timescale of a few weeks \citep{matt}. We thus expect that variability of the maser emission would cause substantial changes to the spectral profiles over ten years, so a detailed comparison of the spectral profile features at $\lambda$3mm and $\lambda$7mm cannot be obtained. 
Nevertheless, broad features of the spectra can still be compared despite the time-variable aspect of the emission. In particular, previous simultaneous observations at $\lambda$3mm and $\lambda$7mm showed differences in the velocity ranges \citep[Matthews et al. in prep]{snyd, doel}. Thus source structure variability is unlikely to be the cause of the differences in the observed velocity ranges between the $\lambda$3mm $v=1$ and the $\lambda$7mm $v=1,2$ transitions. 

\section{Discussion}\label{sec:discussion}
\subsection{The SiO maser kinematics} \label{sec:model}

\begin{figure}
\centering
\includegraphics[width=\linewidth]{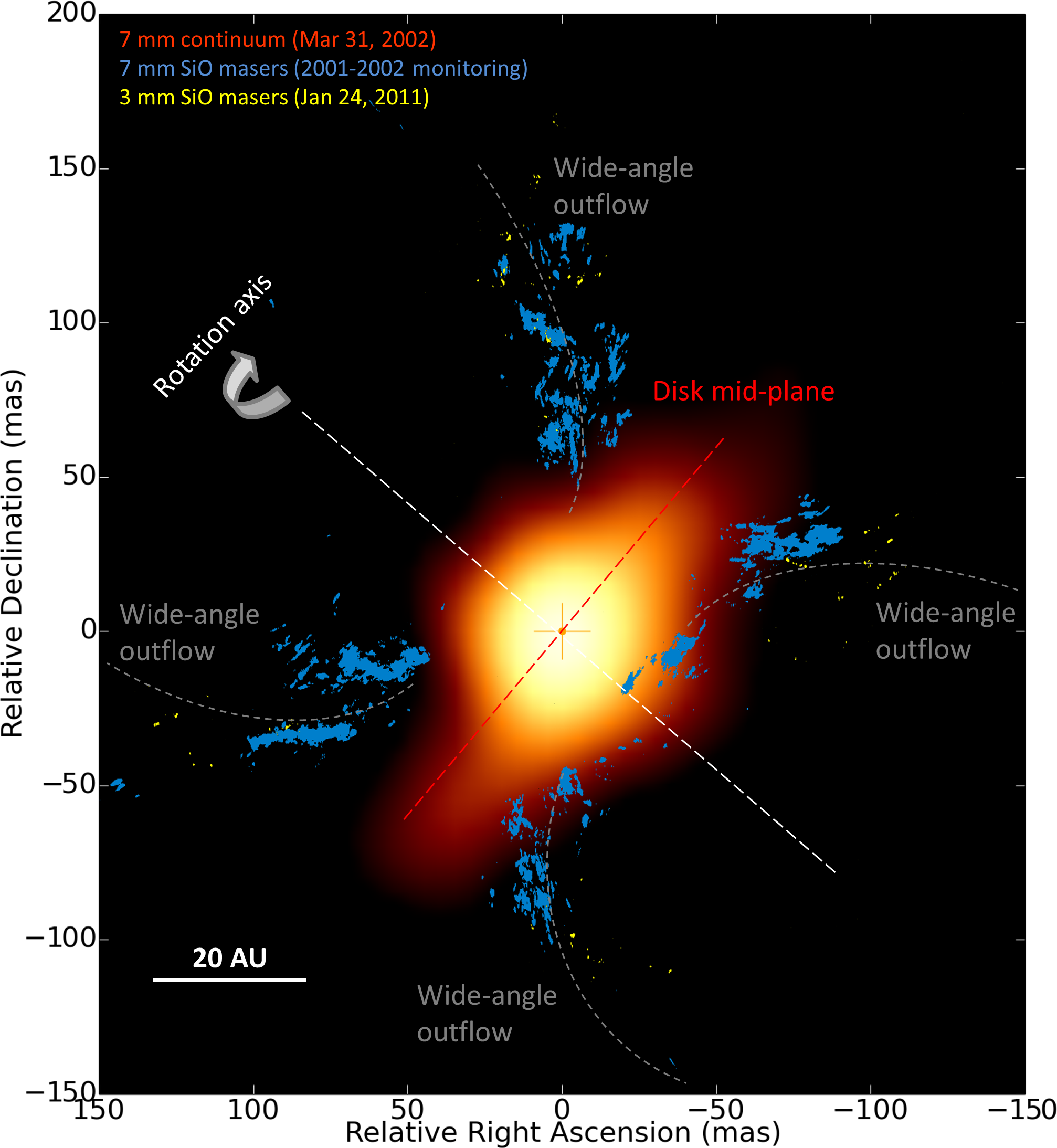}
\caption{Kinematic model for Source~I: the bulk of the $\lambda$3mm and $\lambda$7mm $^{28}$SiO maser emission is located within the four arms of a wide-angle outflow, while a dark band with no SiO emission hosts the $\lambda$7mm continuum source tracing an ionized and dusty disk \citep{reid,plamb}. The putative position of Source~I is assumed to be coincident with the peak of the continuum emission. 
}\label{fig:SourceI}
\end{figure}

Based on the previous observations of the $^{28}$SiO $v=1,2$, $J=1-0$ maser transitions, a disk-wind model has been proposed to explain the gas kinematics around Source~I \citep{greenb, kim, matt}. In this model, illustrated in Fig.~\ref{fig:SourceI}, the arms of SiO emission outline motions of gas clumps along the limbs of a wide-angle outflow arising from the surface of an edge-on disk rotating about a northeast-southwest axis. Our new observations of the $^{28}$SiO $v=1$, $J=2-1$ maser transition emission confirm this model. Specifically, this model is supported by the presence of a continuous velocity gradient in the arms traced by both the $\lambda$3mm and $\lambda$7mm SiO maser emission, which is indicative of a continuous single wide-angle outflow suggesting differential rotation about a northeast-southwest rotation axis. Additionally, the registrations of the $\lambda$3mm and $\lambda$7mm images with respect to the putative location of Source~I exhibit a consistency in the shape and bend of the limbs of the outflow, where the $\lambda$3mm emission appears to fall along the streamlines of the $\lambda$7mm maser emission. 

\subsection{Maser pumping} \label{sec:pump}
Most available pumping models of SiO masers focus on modeling spherically symmetric evolved AGB stars, due to the more common occurrence of these masers around these objects, and typically include both radiative and collisional pumping mechanisms. 
The model by \cite{humphreys} was the first to also include effects from hydrodynamic shocks in stellar envelopes, which may play an important role in the context of the wide-angle outflow model for Source~I. Simulations of the shock regions from an AGB star showed that the $v=1$, $J=2-1$ maser emission occurs at slightly larger radii than the $v=1$, $J=1-0$ masers, with a spatial offset of 2$\%$. This offset, when applied to the case of Source~I, would correspond to a distance shorter than 1\,AU, which does not explain the larger offsets observed in our VLBI images ($\sim$7\,AU on average between the $v=1$ $\lambda$3mm and $\lambda$7mm transitions).

The spatial offset between the $v=1$, $J=1-0$ and $J=2-1$ $^{28}$SiO transitions, although uncommon, has also been observed in evolved stars \citep[e.g.][]{soria}. \cite{soria}, and later \cite{desmurs}, attempted to explain this phenomenon by including line overlap between H$_2$O and SiO in their modeling, which can bring about the $J=2-1$ emission further from the star (by about $50\%$) than the $J=1-0$ emission. In the model by \cite{soria}, the inclusion of the line overlap with H$_2$O resulted in a promising explanation for the maser distributions around the O-rich late-type star IRC+10011: the $v=1$, $J=2-1$ emission lies at radii significantly further away from the star, and the $v=1,2$, $J=1-0$ are spatially coincident close to the star. 
It is worth noting that this addition to the model did not affect significantly the conditions for the masing of the $v=1, J=2-1$ line, but instead shifted the $v=2, J=1-0$ to lower density requirements, where the $v=1,2$, $J=1-0$ lines are now strongly coupled.  

In the case of the model by \cite{desmurs}, a similar mechanism takes place, where the $v=1$, $J=1-0$ emission is shifted to requiring higher densities and thus appears closer to star. Here as well, the $v=1$, $J=2-1$ line is unaffected by the H$_2$O line overlap, and the offset between the two lines in the $v=1$ state is caused by the $J=1-0$ shifting closer to the star, such that the $v=1,2$, $J=1-0$ are co-spatial at small radii from the star and the $v=1$, $J=2-1$ emission lies at larger radii. The inclusion of the line overlap, although creating an offset between the two SiO lines in the $v=1$ vibrational state, fails to explain the observed offset in Source~I between the $v=1,2$, $J=1-0$ lines, where the $v=2$ appears closer (by $\sim10\%$) to the YSO, at higher temperatures and densities. It also fails to explain the partial spatial overlap for both rotational transitions in the $v=1$ vibrational state observed in Source~I.

Another model, by \cite{gray}, includes hot dense dust and can also obtain an offset between the $v=1$, $J=1-0$ and $J=2-1$ emission in an evolved star, without the inclusion of the H$_2$O line overlap. In particular, the $v=1$, $J=2-1$ emission occurs $\sim 25 \%$ further away from the star than the $v=1$, $J=1-0$ emission (comparable to what is observed for Source~I). However, in this particular model, the $v=1,2$, $J=1-0$ transitions are again predicted to be co-spatial and thus sharing the same masing conditions, which is inconsistent with what is observed in Source~I. 

\begin{figure}
\centering
\includegraphics[width=\linewidth]{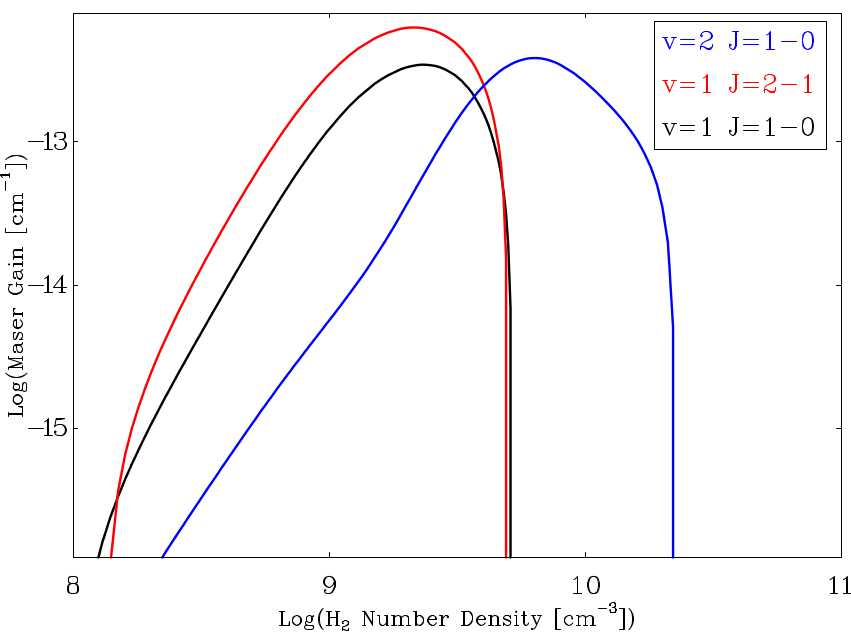}
\caption{Results of the $^{28}$SiO $v=1,2$, $J=1-0$ and $v=1$, $J=2-1$ maser gains as a function of temperature for gas at a kinetic temperature of 2000\,K from the LVG radiative transfer model presented by \cite{goddi}.}\label{fig:LVG}
\end{figure}

From studies of the physical conditions expected for the stimulation of the $v=1,2$, $J=1-0$ transitions from radiative transfer modeling for a single YSO using a large velocity gradient (LVG) code, it was found that the $v=2$ transition can occur at higher densities and temperatures than the $v=1$ transition, while still maintaining a range of densities and temperatures where both transitions can occur \citep[][Fig.~\ref{fig:LVG}]{goddi}. \cite{matt} concluded that the model is consistent with observations of the $v=2$ transition arising closer to the YSO, where higher temperatures and densities are expected, but substantial spatial overlap between the $v=2$ and $v=1$ transitions remained, in accordance with model predictions. 
Similar calculations were done by \cite{cho} for various additional transitions using the same model: most notably, they found that the physical conditions necessary for the $^{28}$SiO $v=1$, $J=2-1$ transition were not readily differentiable from the conditions for the $J=1-0$ transition of the same vibrational state, again in contrast with our findings in Source~I. 

\begin{figure}
\centering
\begin{minipage}{\linewidth}
\includegraphics[width=\linewidth]{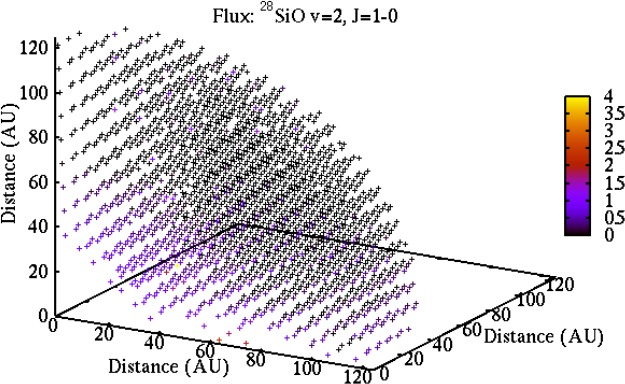}
\end{minipage}
\begin{minipage}{\linewidth}
\includegraphics[width=\linewidth]{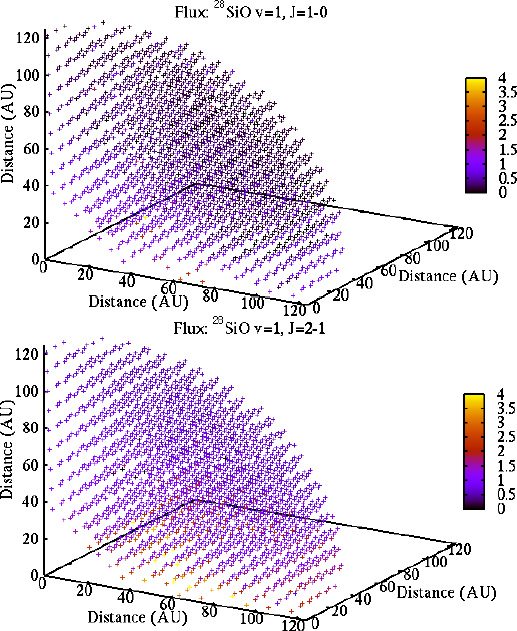}
\end{minipage}
\caption{Intensity distribution of simulated SiO maser clumps as a function of distance from the central binary: \textit{(top)} $^{28}$SiO $v=2$, $J=1-0$ maser clumps; \textit{(center)} $^{28}$SiO $v=1$, $J=1-0$ maser clumps; \textit{(bottom)} $^{28}$SiO $v=1$, $J=2-1$ maser clumps.}\label{fig:Malcolm}
\end{figure}

\begin{table*}[!ht]
\caption{Summary of the main features of relevant SiO maser pumping models currently available and their predicted relative spatial offsets between the three transitions.}\label{tab:models}\centering 
\small 
\begin{tabularx}{\linewidth}{llXXX}
\hline
\hline 
\noalign{\smallskip}
Model & Type of system & Main feature  & $v=2$, $J=1-0$ with respect to $v=1$, $J=1-0$  & $v=1$, $J=2-1$ with respect to $v=1$, $J=1-0$ \\
\noalign{\smallskip}
\hline
\noalign{\smallskip}
\cite{humphreys} & single evolved star & hydrodynamic shocks & $\sim 0\%$ (co-spatial) & $\sim 2\%$ further away \\

\cite{soria} & single evolved star & overlaps with H$_2$O absorption lines & $\sim 0\%$ (co-spatial) & $\sim 50\%$ further away \\

\cite{gray}  & single evolved star & hot dense dust & $\sim 0\%$ (co-spatial) & $\sim 25\%$ further away \\

\cite{goddi} & single YSO & radiative transfer (LVG) & slightly closer in & $\sim 0\%$ (co-spatial) \\

Gray et al. (in prep)  & binary YSO & full hydrodynamic solution for binary & $\sim 3\%$ closer in & $\sim 20\%$ further away \\
\noalign{\smallskip}
\hline 
\end{tabularx}
\end{table*}

\begin{figure}
\centering
\includegraphics[width=\linewidth]{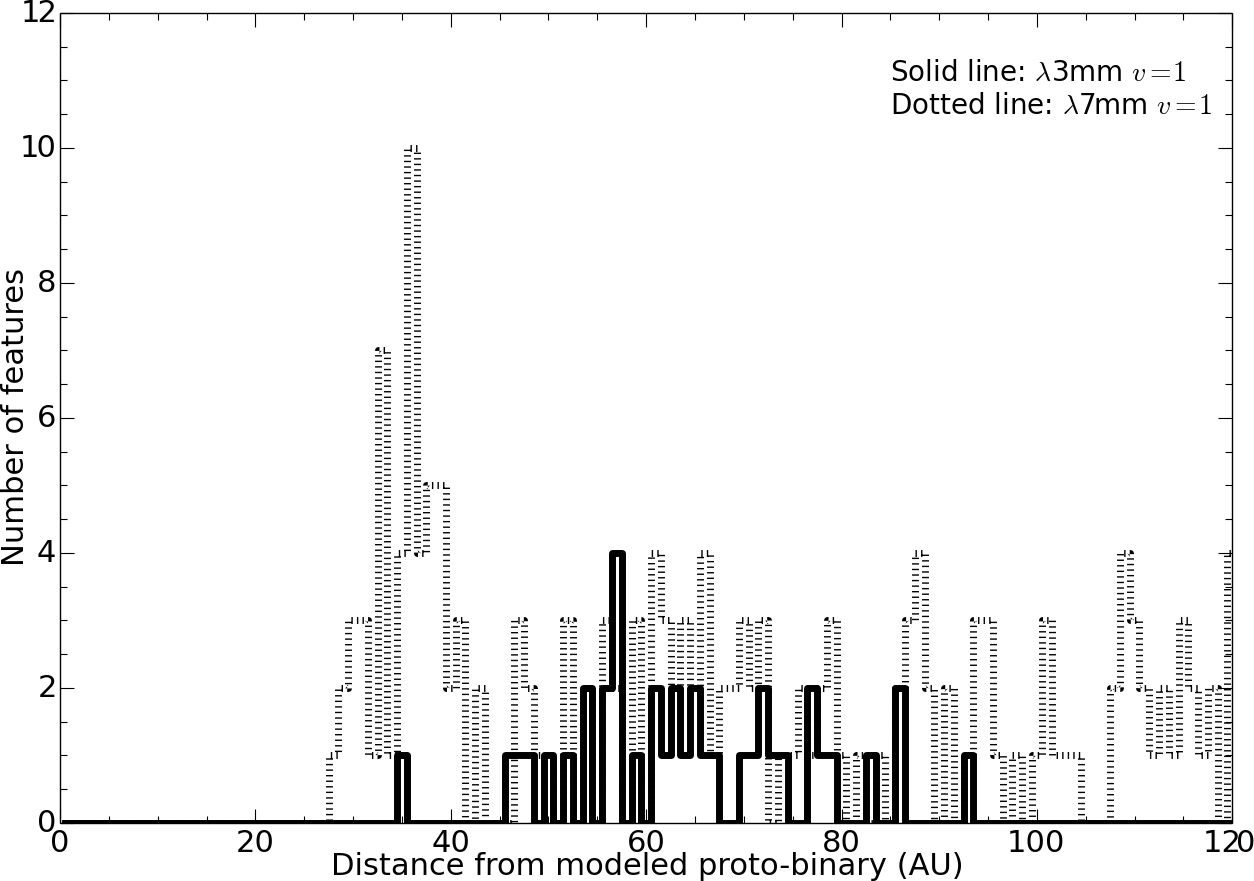}
\caption{Histograms showing the number of individual maser spots selected in the $\protect\lambda$3mm $v=1$, $J=2-1$ (solid line) and $\protect\lambda$7mm $v=1$, $J=1-0$ (dotted line) emission and their positions relative to the central object in the binary YSO pumping model (Gray et al. in prep). The histogram bin width is 1\,AU.}\label{fig:spots1}
\end{figure}

\begin{figure}
\centering
\includegraphics[width=\linewidth]{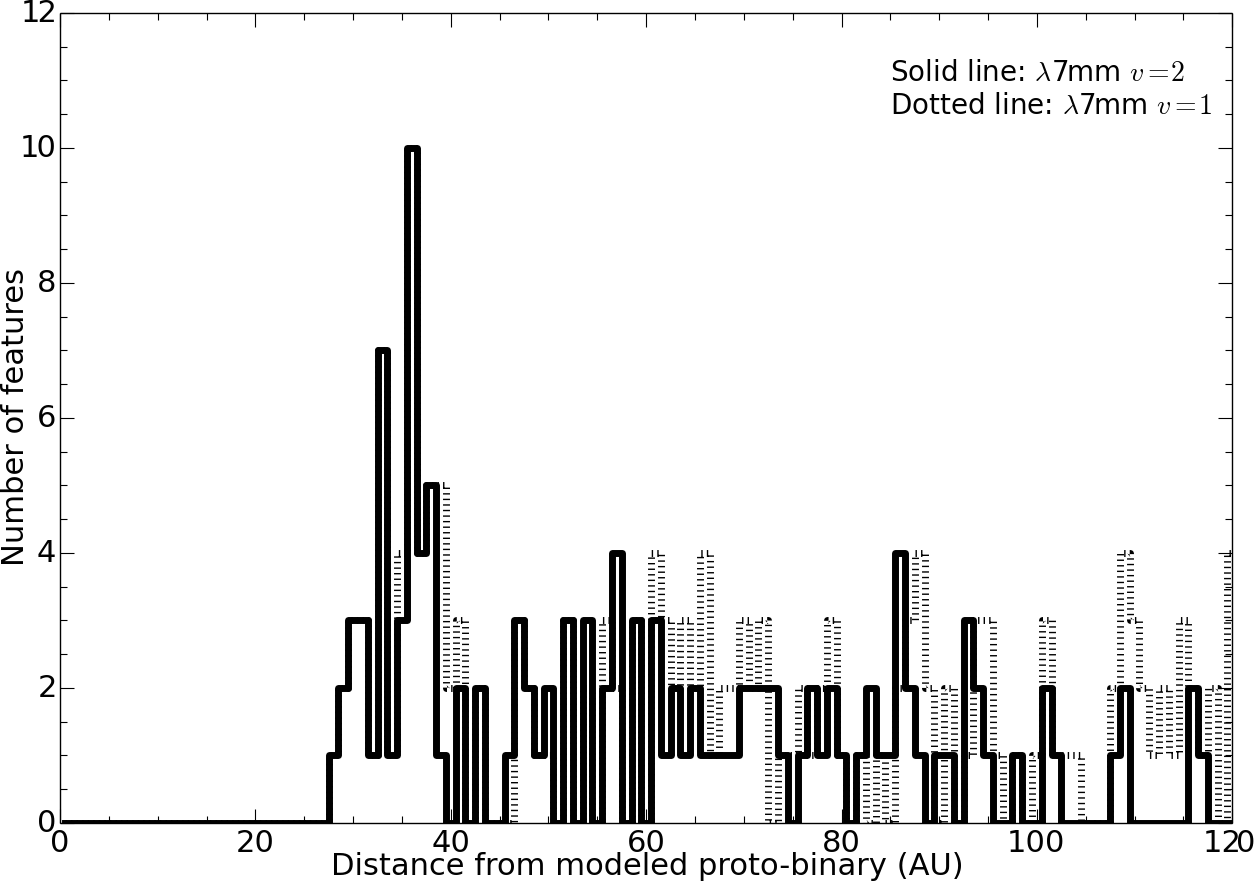}
\caption{Histograms showing the number of individual maser spots selected in the $\protect\lambda$7mm $v=1$, $J=1-0$ (dotted line) and $v=2$, $J=1-0$ (solid line) emission and their positions relative to the central object in the binary YSO pumping model (Gray et al. in prep). The histogram bin width is 1\,AU.}\label{fig:spots2}
\end{figure}

A new model by Gray et al. (in prep) investigates the pumping conditions of SiO masers arising in the disk-midplane of a close massive proto-binary surrounded by a circumbinary disk, using a full hydrodynamic solution by \cite{krum}. The model uses an accelerated lambda iteration (ALI) radiative transfer model and includes all SiO isotopologues ($^{28}$SiO, $^{29}$SiO, $^{30}$SiO) with full line overlap. The masers are represented by point-like ``bullet" clouds (as shown in Fig.~\ref{fig:Malcolm}), with positions and physical conditions specified by the grid-nodes within the hydrodynamic model \citep[as in Fig.~1 of ][]{krum9}. A more detailed description will be given in Gray et al. (in prep). This model is potentially more suitable for the case of Source~I, which is believed to be a massive proto-binary \citep{goddi2}. The resulting physical conditions for the maser transitions differ from those of an evolved single star: the $v=1$, $J=2-1$ transition occurs at slightly lower densities: $J=2-1$ at a mean density of $\sim 3.1 \times 10^8$\,cm$^{-3}$ and $v=1,2$ $J=1-0$ at $\sim 3.5 \times10^8$\,cm$^{-3}$ and $\sim4.7 \times 10^8$\,cm$^{-3}$ respectively.   

We constructed histograms of the maser spot distributions as a function of distance from the proto-binary as predicted by the new pumping model in order to compare to the observed spatial distributions for the three maser transitions. Since the model gives a spherically symmetric distribution of maser spots, the number of spots increases with radius (Fig.~\ref{fig:Malcolm}). Therefore, in constructing the histograms, we only selected the maser spots of reasonable brightness, within a conservative range of 500--1000 times the brightness of the weakest spots in the distributions. The resulting histograms comparing the $v=1$, $J=1-0$ and $J=2-1$ and the $v=1,2$, $J=1-0$ distributions are presented in Fig.~\ref{fig:spots1} and Fig.~\ref{fig:spots2}, respectively. 

The mean radii (computed from the histograms) for the $v=1,2$, $J=1-0$ and $v=1$, $J=2-1$ are 69\,AU, 61\,AU and 65\,AU respectively. This would imply that the $v=1$, $J=2-1$ transition lies at closer ($\sim 6\%$) projected distances from the binary than the $v=1$, $J=1-0$ transition (Fig.~\ref{fig:spots1}), in contrast with our findings in Source~I. However, this offset is computed over the full extent of the hydrodynamic solution (up to 120\,AU), while we know that no vibrationally-excited maser emission is observed beyond 80\,AU in Source~I, where it is effectively quenched by the $v=0$ $^{28}$SiO emission \citep{green}. 
In fact, when we apply this observational constraint to the modeled maser distributions, i.e. truncating all maser spots appearing beyond 80\,AU, the mean radii for the $v=1,2$, $J=1-0$ and the $v=1$, $J=2-1$ are reduced to 52\,AU, 50\,AU and 62\,AU respectively\footnote{The truncation of maser spots primarily affects the $v=1,2$, $J=1-0$ distributions, which have tails of features at radii 80--120\,AU from the central object (see Figs.~\ref{fig:spots1} and \ref{fig:spots2})}. As a result of this truncation, the $\lambda$3mm $v=1$ masers arise $\sim 20\%$ further away, on average, from the central object than the $v=1$ $\lambda$7mm masers, reproducing the observed offset in our VLBI images. In addition, the $v=1,2$ $\lambda$7mm masers have a $\sim 3\%$ offset, which is smaller than (but in the same direction of) that observed in Source~I ($\sim 10\%$). 

In summary, this new pumping model is the only one (to date) able to reproduce the observed $v=1$ $\lambda$3mm and $\lambda$7mm offset as well as (to a lesser extent) the $v=1,2$ $\lambda$7mm offset (Table~\ref{tab:models}). 
Some caveats are however in order. First, the model by \cite{krum} describes a binary composed of an 8.3\,M$_\odot$ primary and a much smaller secondary, whereas Source~I is believed to be an equal-mass binary of a total mass of 20\,M$_\odot$ \citep{goddi2}. The ratio between the primary and secondary masses is likely to affect the system as a whole and thus modify the pumping conditions of SiO masers arising from the circumbinary disk. Second, the model does not specify an SiO abundance and does not include a dust model, shown by \citet{gray} to affect the masing conditions, or effects from line overlap with the H$_2$O absorption lines \citep{soria, desmurs}, which could potentially explain our findings in Source~I. Third, this model is first and foremost a pumping code for evolved stars adapted for star formation, and thus is not expected to accurately explain observations in a high-mass star formation environment, where strong UV radiation affects the surrounding medium. 

\section{Summary and Conclusions}\label{sec:summary}
We have presented the first images of the $^{28}$SiO $v=1$, $J=2-1$ maser emission around Source~I observed with the VLBA. We found that the $\lambda$3mm masers lie in an X-shaped locus consisting of four arms, with blue-shifted emission in the south and east arms and red-shifted emission in the north and west arms. Comparisons with previous images of the $v=1,2$, $J=1-0$ transitions, observed by \cite{matt} in 2001--2002, showed that the bulk of the $J=2-1$ transition emission follows the streamlines of the $J=1-0$ emission and exhibits an overall velocity gradient consistent with the gradient at $\lambda$7mm. We interpret the $\lambda$3mm and $\lambda$7mm masers as being part of a single wide-angle outflow arising from the surface of an edge-on disk rotating about a northeast-southwest rotation axis, with a continuous velocity gradient indicative of differential rotation of the gas, consistent with a Keplerian profile.

We find that the $v=1$, $J=1-0$ and $J=2-1$ transitions share spatial overlap, suggestive of a common range of temperatures and densities where physical conditions are favorable for both rotational transitions of a same vibrational state. Despite this overlap, we also find that the bulk of emission at $\lambda$3mm arises significantly further away from the YSO than the $\lambda$7mm emission (average offset $\sim$9\,AU). The observed spatial offset indicates different ranges of temperatures and densities for optimal excitation of the different rotational maser transitions.

We have discussed several existing pumping models for SiO masers, which however fail in reproducing the observed offsets. This is not a surprising result, as most available pumping models of SiO masers are built for spherically symmetric evolved AGB stars (where SiO masers commonly occur), and are not expected to faithfully describe a high-mass star formation environment. 
We present here a promising new model by Gray et al. (in prep) using a full hydrodynamic solution of a close massive proto-binary \citep{krum}, to assess the pumping conditions of SiO masers in the circumbinary disk-midplane. Given the binary YSO nature of Source~I \citep{goddi2}, this model is potentially more suitable than AGB models. In fact, we were able to reproduce the offset between the $v=1$ $\lambda$3mm and $\lambda$7mm masers observed in Source~I ($\sim 20\%$) and an additional offset ($\sim 3\%$) for the $v=1,2$ $\lambda$7mm masers, although smaller than found in Source~I.
 
Future models may need to include additional physical mechanisms known to be at work in high-mass star-forming environments such as the effects of a disk-wind, uv-radiation, and/or magnetic fields, in order to reproduce and explain the spatial offsets for all three maser transitions simultaneously. 
In addition, owing to variability of the SiO maser emission, simultaneous VLBI observations of the $\lambda$3mm and $\lambda$7mm transitions will be required for a more detailed comparative study of the maser structures and kinematics between these transitions.

\begin{acknowledgements} 
The data presented here were part of the NRAO program BG205.
\end{acknowledgements} 

\bibliographystyle{aa}
\bibliography{Sio_bib}

\end{document}